\documentclass[twocolumn,fp]{jpsj3}
\usepackage{graphicx,exscale,amsmath,amssymb}
\usepackage{mathrsfs}
\usepackage{ascmac}
\def\v#1{\mib #1}

\def\Ms{M_{\rm s}}
\def\SF{{SF}}
\def\LD{{LD}}
\def\PDN{{PDN}}
\def\PF{{PF}}
\def\QF{{QF}}
\def\sztot{M}

\newcommand{\ket}[1]{\left\vert {#1} \right\rangle}

\newcommand{\aver}[1]{\left\langle {#1} \right\rangle}

\title
{
Ground-State Phase Diagram of $S=2$  Heisenberg Chains with  Alternating Single-Site Anisotropy }

\author
{
Kazuo { Hida}\thanks{E-mail address: hida@mail.saitama-u.ac.jp}
}

\inst
{Division of Material Science, Graduate School of Science and Engineering,\\ Saitama University, Saitama, Saitama 338-8570, Japan
 
}

\recdate
{
November 1, 2013
}

\abst
{
The ground-state phase diagram of $S=2$ antiferromagnetic Heisenberg chains with coexisting uniform and alternating single-site anisotropies is investigated by the numerical exact diagonalization and density matrix renormalization group methods. We find the Haldane, large-$D$, N\'eel, period-doubled N\'eel, gapless spin fluid, quantized and partial ferrimagnetic phases. The Haldane phase is limited to the close neighborhood of the isotropic point. Within numerical accuracy, the transition from the gapless spin-fluid phase to the period-doubled N\'eel phase  is a direct transition. 
 Nevertheless, the presence of a narrow spin-gap phase between these two phases 
 is suggested on the basis of the low-energy effective theory. The ferrimagnetic ground state is present in a wide parameter range. This suggests the realization of magnetized single-chain magnets with a uniform spin magnitude by controlling the environment of each magnetic ion without introducing ferromagnetic interactions.

}

\begin{document}
\sloppy
\maketitle
\section{Introduction}

Among various exotic ground states in quantum magnetism, the Haldane state in  integer spin antiferromagnetic Heisenberg chains\cite{fd} has been most extensively studied both experimentally and theoretically. In the case of $S=1$, this state is characterized by a hidden antiferromagnetic string order accompanied by the $Z_2\times Z_2$ symmetry breakdown in spite of the presence of the energy gap and the exponential decay of the spin-spin correlation function. The easy-plane single-site anisotropy $D (>0)$ destroys the Haldane phase, leading to the large-$D$ ({\LD}) phase with a finite energy gap and an exponentially decaying spin-spin correlation function without the string order. The easy-axis  single-site anisotropy ($D <0$) drives the Haldane state into the N\'eel state.\cite{md,ht} On the other hand, in the case of $S=2$, the Haldane and {\LD} phases belong to the same topological phase, as pointed out by Pollmann {\it et al.}\cite{pollmann2012} and Tonegawa {\it et al.}\cite{tonegawa2011} 

In this context, it is interesting to investigate how the ground states of the quantum spin chains are modified if the  easy-axis and easy-plane single-site anisotropies coexist in a single chain. In a previous work,\cite{altd} the present author and Chen  investigated the $S=1$ chain with  coexisting uniform ($D_0$) and alternating  ($\pm\delta D$)  single-site anisotropies and found the period-doubled N\'eel ({\PDN}) phase with a $\ket{\uparrow 0 \downarrow 0}$ structure for large values of $\delta D$. In this model, the Haldane phase is stable as long as one of the single-site anisotropies $D_0\pm \delta D$ is not much larger than the exchange coupling $J$. 

  The present author also investigated this problem in the case of $S=2$ with only the alternating single-site anisotropy $\pm \delta D$.\cite{kh} We found not only the nonmagnetic and period-doubled N\'eel phase, but also the ferrimagnetic phases with quantized and unquantized spontaneous magnetization for intermediate values of $\delta D$. These quantized values of magnetization also satisfy the Oshikawa-Yamanaka-Affleck condition\cite{oya}, well-known for the magnetization plateau in the magnetic field. However, this model is rather special, because the magnitudes of the easy-axis and easy-plane anisotropies are assumed to be equal. In the present work, we investigate the case of $S=2$ in a more general situation with coexisting uniform and alternating  single-site anisotropies. A similar alternation in $D$ is experimentally realized in a mixed spin molecular magnet in which the magnitude of spins also alternates between $S=1$ and 2.\cite{MnNi}  This type of model has  also been investigated theoretically.\cite{tonemix} In the present work, we fix $S=2$ for all sites to single out the effect of alternation in the single-site anisotropy.

This paper is organized as follows. In the next section, the model Hamiltonian is presented.  In sect. 3, the numerical results for the ground-state phase diagram are presented and the property of each phase is discussed. Details of the numerical analysis are explained in sect. 4. The last section is devoted to a summary and discussion.

\section{Model Hamiltonian}

We investigate the ground state of the $S=2$ antiferromagnetic Heisenberg chains with the alternating single-site anisotropy described by the Hamiltonian
\begin{eqnarray}
\label{ham0}
{\cal H} &=& \sum_{l=1}^{N}J\v{S}_{l}\v{S}_{l+1}+(D_0+\delta D)\sum_{l=1}^{N/2}S_{2l-1}^{z2}\nonumber\\
&+&(D_0-\delta D)\sum_{l=1}^{N/2}S_{2l}^{z2},
\end{eqnarray}
where  $\v {S_{i}}$ is the $S=2$ spin operator on the $i$-th site. We consider the antiferromagnetic case $J > 0$.  We also take  $\delta D >0$ without the loss of generality. We investigate this model using the numerical exact diagonalization (NED) and density matrix renormalization group (DMRG) methods.

\begin{figure}[h]
\centerline{\includegraphics[width=70mm]{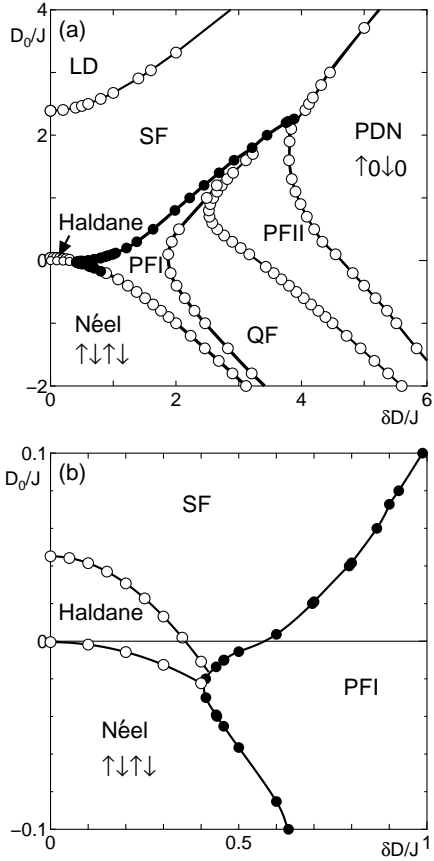}}
\caption{Ground-state phase diagram. 
 Open symbols are determined by extrapolation from the NED data for $4 \leq N \leq 12$. Filled symbols are determined from DMRG data with $N=60$. (a) Overall phase diagram. (b) Enlarged phase diagram around $(D_0, \delta D) \sim (0,0)$.}
\label{phase}
\end{figure}
\section{Ground-State Phase Diagram}
Our results are summarized in the phase diagram of Fig. \ref{phase}. 
As in the case of the uniform $D$, the Haldane phase is fragile against anisotropy. It occupies  
only a small region around the isotropic point $(D_0, \delta D)= 0$, as shown in Fig. \ref{phase}(b).  Instead of the Haldane phase, a wide gapless spin-fluid (SF) phase and a conventional N\'eel-ordered phase are realized for $D_0 >0$ and  $D_0 < 0$, respectively. 

 For a large positive $D_0$, all spins are confined in the state  $S^z_i=0$. As a result, 
 a phase transition to the {\LD} phase occurs with an increase in $D_0$. As pointed out by Pollmann {\it et al.}\cite{pollmann2012} and Tonegawa {\it et al.},\cite{tonegawa2011} the Haldane and {\LD} phases of the $S=2$ chain belong to the same topological phase. Within the present model, however, we find no direct continuous path to connect these two phases. 

The natures of the Haldane, LD, and N\'eel phases are essentially the same as those in the uniform case. In the following subsections, we explain the natures of the SF, PDN, and ferrimagnetic phases separately in detail. 
\subsection{SF phase}
{Even for a large positive $D_0$,  the easy-plane anisotropy decreases on the $2l$-th site with an increase in $\delta D$, and the gapless {\SF} phase is recovered. To obtain insight into the nature of the SF phase in this regime, we examine the limit $D_0+\delta D \gg J,\ | D_0-\delta D|$. In this limit, only the spins $\v{S}_{2l}$ survive. The effective Hamiltonian for these spins is obtained by the second-order perturbation in $J$ as
\begin{align}
{\cal H}_{\rm eff} &= \sum_{l=1}^{N/2}\Big[J_{\rm eff}^{xy}\left({S}^x_{2l}{S}^x_{2(l+1)}+{S}^y_{2l}{S}^y_{2(l+1)}\right)\nonumber\\
&
+(D_0-\delta D)S^{z2}_{2l}\Big],\label{hameff}
\end{align}
where $J_{\rm eff}^{xy}=-6J^2/(D_0+\delta D)$. Hence, the effective model is equivalent to the $S=2$ XY chain with a single-ion anisotropy up to the second order in $J$. 
The ferromagnetic XY coupling implies that the ferromagnetic quasi-long-range order in the $x$- and $y$-components of the spins $\v{S}_{2l}$ is dominant in the SF phase. Hence, the correlation functions should behave as $\aver{{S}^x_{2l}{S}^x_{2(l+j)}}=\aver{{S}^y_{2l}{S}^y_{2(l+j)}}  \sim j^{-\eta(D_0,\delta D)}$.  On the other hand, in the SF phase with $\delta D=0$, the dominant correlation is antiferromagnetic in the $x$- and $y$-components of all spins as $\aver{{S}^x_{l}{S}^x_{l+j}}=\aver{{S}^y_{l}{S}^y_{l+j}} \sim (-1)^jj^{-\eta(D_0,0)}$.  In the presence of a finite $\delta D$, however, the unit cell is doubled. In this case, the correlation should be calculated between equivalent sites in different unit cells. To compare the correlations for the general values of $D_0$ and $\delta D$ with those in the limiting case of  $D_0+\delta D \gg J,\ | D_0-\delta D|$, we concentrate on the correlations between the spins $\v{S}_{2l}$. For a small $\delta D$, they should  behave as $\aver{{S}^x_{2l}{S}^x_{2(l+j)}}=\aver{{S}^y_{2l}{S}^y_{2(l+j)}} \sim j^{-\eta(D_0,\delta D)}$, taking into account the continuity to the case of $\delta D=0$. Thus, the dominant correlation is common in both limiting cases. This observation suggests that the SF phase is a single phase irrespective of $D_0$ and $\delta D$. We have also confirmed that the correlations $\aver{{S}^x_{2l}{S}^x_{2(l+j)}}$ and $\aver{{S}^y_{2l}{S}^y_{2(l+j)}}$ are always ferromagnetic within the SF phase by  NED calculation for $N=12$.} 

\subsection{PDN phase}
With further increase in $\delta D$, the anisotropy on the $2l$-th site changes into the easy-axis type and a phase transition to the {\PDN} phase 
 that has a spin configuration $\uparrow 0\downarrow 0$ occurs. In the limit of $\delta D \gg D_0,\ J$, the origin of the effective antiferromagnetic Ising interaction between $S^z_{2l}$ and $S^z_{2l+2}$ can be understood in the same way as in the case of $S=1$.\cite{altd} Let us denote the spin state of three successive sites of the Hamiltonian (\ref{ham0}) by $\ket{S^z_{2l}\ S^z_{2l+1}\ S^z_{2l+2}}$. In the absence of the exchange term $J$, the ground states are $\ket{\pm 2\ 0\ \pm 2}$, which are fourfold degenerate. Owing to the spin flip term in (\ref{ham0}), the $\ket{2\ 0\ 2}$ state is mixed with the $\ket{2\ 1\ 1}$ and  $\ket{1\ 1\ 2}$ states, whereas the $\ket{2\ 0\ {-2}}$ state is mixed with the $\ket{2\  {-1}\ -1}$ and  $\ket{1\ 1\ {-2}}$ states. Hence, the energy of the $\ket{2\ 0\ 2}$ state is raised, whereas that of the $\ket{2\ 0\ {-2}}$ state is lowered by the Ising component of the nearest-neighbour coupling. This means that the effective interaction between $S^z_{2l}$ and $S^z_{2l+2}$ is antiferromagnetic. 

 If the SF-PDN transition is a single transition, it  corresponds to the Brezinskii-Kosterlitz-Thouless (BKT) transition  with a $Z_2$ symmetry breakdown. 
 However, it is possible that the BKT transition and the Ising-type transition with a $Z_2$ symmetry breakdown take place separately with a narrow intermediate spin-gap phase in between. This issue will be examined in detail in sect. \ref{section:sf_dble}.  
\subsection{Ferrimagnetic phases}

For $-\infty < D_{0} < D_{\rm 0c} \simeq 2.26 J$,  
ferrimagnetic phases also appear 
 in a wide parameter range, as in the case of $D_0=0$.\cite{kh}    
In the middle of the ferrimagnetic phase, there exists a quantized ferrimagnetic (QF) phase, where the total magnetization $M$
 is quantized to $M=\Ms/4$, except in the regime $D_{0} \lesssim D_{\rm 0c}$. Here, $\Ms=2N$ is the saturation magnetization. This value of $M$ satisfies the quantization condition by Oshikawa {\it et al.}
\cite{oya}  given by
\begin{equation}
p(S-m)=q,
\label{oymcond}
\end{equation}
where $p$ is the size of the unit cell, $q$ is an integer, and $m$ is the magnetization per site ($m=M/N=MS/\Ms$). Although this condition is originally proposed for the magnetization plateau in the magnetic field, it also holds for the spontaneous magnetization in the ferrimagnetic phase.\cite{kh} The present QF state corresponds to the case of $p=2$, $S=2$, $m=1/2$, and $q=3$ in (\ref{oymcond}). 

 On  both sides of the quantized ferrimagnetic phase, we find two kinds of partial ferrimagnetic (PF) phases, namely, the PFI and PFII phases.   
In the {\PF}I phase, the spontaneous magnetization varies continuously from 0 at the {\PF}I-{\SF} phase boundary to $\Ms/4$ at the  {\PF}I-{\QF} phase boundary.  In the {\PF}II phase, it  varies continuously from $\Ms/4$ at the {\QF}-{\PF}II phase boundary to a critical value $M_{\rm cr}$ at the PFII-PDN phase boundary. The critical value $M_{\rm cr}$ varies from point to point on the PFII-PDN boundary. 
The properties of the ferrimagnetic phases are qualitatively the same as those in the case of $D_0=0$, which have been discussed in detail in Ref. \citen{kh}.

\section{Numerical Determination of Phase Boundaries}
\subsection{Phase boundary between the phases with different spontaneous magnetization}

The ground-state spontaneous magnetization is calculated by NED with a periodic boundary condition and by DMRG with an open boundary condition for various values of $D_0$ and $\delta D$.  
For the DMRG calculation, appropriate end spins are added to reduce the boundary effects.

The phase boundaries between the ground states with different values of spontaneous magnetization are determined by the level crossing among them. The NED calculation is carried out for $N=4, 8$, and 12. For small $D_0$ and $\delta D$, however, the phase boundary between the {\SF} and {\PF}I phases, and that between the N\'eel and {\PF}I phases strongly depends on the system size. Hence, we employ the DMRG calculation for $N=60$ with an open boundary condition. For large values of $D_0$, the phase boundary thus obtained almost coincides with that obtained by NED with a periodic boundary condition  for $N=12$. 
\subsection{N\'eel-Haldane phase boundary}
\label{subsec:nh}
\begin{figure}
\centerline{\includegraphics[width=70mm]{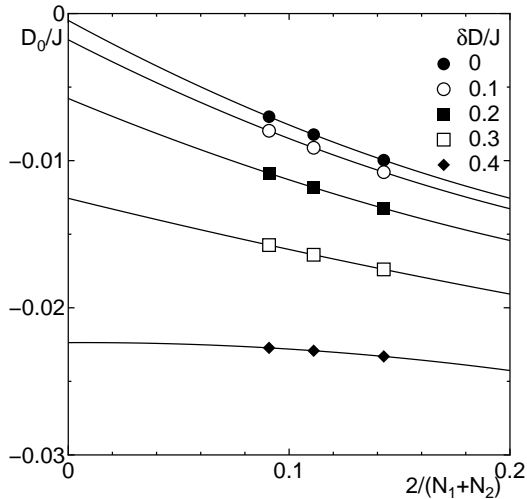}}
\caption{Extrapolation procedure for the N\'eel-Haldane critical points using $(N_1,N_2)=(6, 8), (8,10)$, and (10,12).}
\label{hn_extra}
\end{figure}

 In the N\'eel phase, the $Z_2$ symmetry is spontaneously broken. This implies that the ground state of a large but finite chain is an antisymmetric superposition of two ordered states with $\sztot=0$ that are interchanged with each other by spin inversion. The lowest excited state is  their symmetric superposition that degenerates with the ground state in the thermodynamic limit. Therefore, the lowest excitation gap with $\sztot=0$ exponentially decreases with the system size. In the Haldane phase, all excitations are gapped. Hence, the phase boundary is determined by  phenomenological renormalization group analysis for the excitation energy $\Delta E(N,\sztot=0)$ with the total magnetization $\sztot=0$. The finite size critical point is determined by the condition
\begin{align}
N_1\Delta E(N_1,\sztot=0)=N_2\Delta E(N_2,\sztot=0),
\end{align}
for $(N_1,N_2)=(6, 8), (8,10)$, and (10,12). The extrapolation to the thermodynamic limit is carried out assuming the principal size dependence $2/(N_1+N_2)$ based on the Ising  exponent  $\nu=1$ for the correlation length, as shown in Fig. \ref{hn_extra}. For $N=8$ and 12, the number of unit cells is even, whereas it is odd for $N=6$ and 10. Therefore, we are concerned about an even-odd oscillation in the extrapolation procedure. Fortunately, for the present transition, no distinct oscillation is observed. 
\subsection{{\LD}-{\SF} and Haldane-{\SF} phase boundaries}
These transitions are conventional BKT transitions. Therefore, we employ the level spectroscopy method with a twisted boundary condition proposed by Nomura and Kitazawa.\cite{nomura-kitazawa} The finite-size critical point is determined by
\begin{align}
E_{\rm TW}(N, \sztot=0)=E(N,\sztot=2),
\end{align}
where  $E(N,\sztot)$ and $E_{\rm TW}(N, \sztot)$ are the ground-state energies in the sector with the magnetization $\sztot$ under periodic and twisted boundary conditions, respectively.

\begin{figure}
\centerline{\includegraphics[width=70mm]{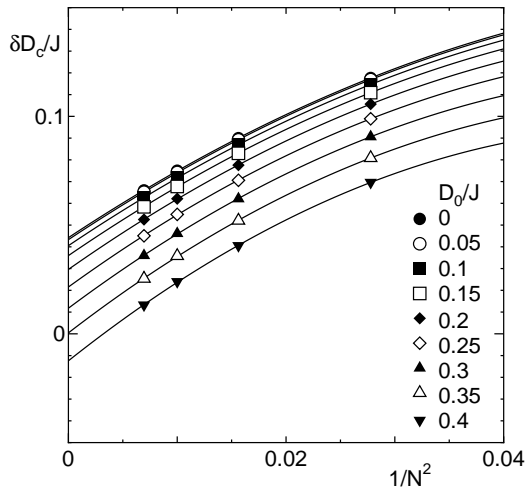}}
\caption{Extrapolation procedure for the Haldane-{\SF} critical points using $N=6, 8, 10$, and 12.}
\label{hf_extra}
\end{figure}

For the {\LD}-{\SF} transition, the extrapolation to the thermodynamic limit  is carried out using the data for $N=8,10$, and 12. In this case, the critical point is insensitive to the system size. 
For the Haldane-{\SF} transition, the extrapolation to the thermodynamic limit  is carried out as shown in Fig. \ref{hf_extra} using the data for $N=6, 8, 10$, and 12, assuming that the finite size correction is $O(N^{-2})$.\cite{nomura-kitazawa} In this case, the even-odd oscillation is also not harmful to the extrapolation.

\subsection{{\SF}-{\PDN} boundary}
\label{section:sf_dble}
\begin{figure}
\centerline{\includegraphics[width=70mm]{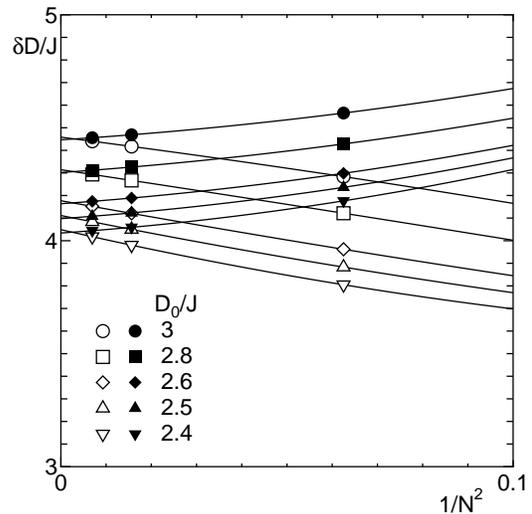}}
\caption{Extrapolation procedure for the critical points of {\SF}-{\PDN} direct transition using $N=4, 8$, and 12 (open symbols). The filled symbols are the estimates obtained assuming the {\SF}-spin-gap BKT transition.}
\label{xy_neel_bkt_extra}
\end{figure}
If this transition is a single transition, it corresponds to the BKT transition with a $Z_2$ symmetry breakdown. The level spectroscopy method for this type of transition was proposed by Okamoto  and Nomura.\cite{okamoto-nomura,nomura-okamoto,nomura-okamoto_JPA} In the {\PDN} phase, the $Z_2$ symmetry is spontaneously broken. 
Hence, the lowest excited state has $\sztot=0$, as in the case of the conventional N\'eel phase discussed in sect. \ref{subsec:nh}. In the gapless {\SF} phase, the lowest excitation is the spin wave with $\sztot=1$. As a result,  the excitation energy  with $\sztot=0$ and that with $\sztot=1$ degenerate at the phase boundary. This also implies that the low-energy excitation spectrum recovers its SU(2) symmetry at the phase boundary in spite of the apparent absence of the SU(2) symmetry in the Hamiltonian (\ref{ham0}). The extrapolation procedures for the critical value of $\delta D$ for different values of $D_0$ are shown in Fig. \ref{xy_neel_bkt_extra} by open symbols. The phase boundary thus obtained is plotted in Fig. \ref{phase}.

\begin{figure}
\centerline{\includegraphics[width=70mm]{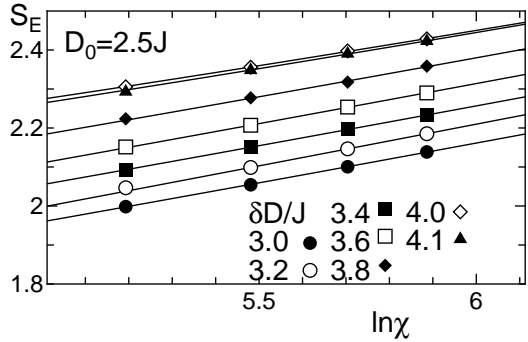}}
\caption{$\chi$ dependence of entanglement entropy $S_{\rm E}$ for $D_0=2.5J$ and various values of $\delta D$ within the {\SF} phase.}
\label{fig:entro}
\end{figure}

However, it is more natural to assume that the BKT transition and the Ising-type transition with a $Z_2$ symmetry breakdown take place separately with a narrow spin-gap phase between them. To obtain more insight into this point, we estimate the central charge $c$ in the {\SF} phase from the behavior of the entanglement entropy $S_{\rm E}(\chi)$ of the optimized matrix product state with a dimension $\chi$.\cite{pollmann2009} It is given by
\begin{align}
S_{\rm E}(\chi)&=\frac{c}{6}\ln \xi_{\chi},
\end{align}
where $\xi_{\chi}$ is the correlation length of the matrix product state that behaves as
\begin{align}
\xi_{\chi} &\propto \chi^{\kappa},
\end{align}
for $\chi \rightarrow \infty$. The exponent $\kappa$ is related to the central charge as
\begin{align}
\kappa&=\frac{6}{c}\frac{1}{\sqrt{\frac{12}{c}+1}}.
\end{align}
Hence, the $\chi$ dependence of $S_{\rm E}(\chi)$ is expected to behave as
\begin{align}
S_{\rm E}(\chi)&\simeq\frac{1}{\sqrt{\frac{12}{c}+1}}\ln\chi+{C_0}+\frac{C_1}{\ln\chi},\label{eq:fit}
\end{align}
for $\chi \rightarrow \infty$, where $C_0$ and $C_1$ are constants. We have estimated $S_{\rm E}(\chi)$ by the infinite-size DMRG method. The $\chi$ dependence of $S_{\rm E}(\chi)$  is consistent with  (\ref{eq:fit}), assuming $c=1$ within the {\SF} phase, as shown in Fig. \ref{fig:entro}. 

Therefore, we  may assume that the low-energy effective Hamiltonian in the SF phase is a Gaussian model given by 
\begin{align}
\label{eq:gauss}
{\cal H}_{0}&= \frac{1}{2\pi} \int {dx \Big[ v_{\rm {s}}K (\pi\Pi)^{2}+\frac{v_{\rm{s}}}{K}\left(\frac{\partial \phi}{\partial x}\right)^{2} \Big]}, \ \ (0 \leq \phi < \frac{2\pi}{\sqrt{2}}),
\end{align}
where $\phi(x)$ is a bosonic field and $\Pi(x)$ is the momentum density conjugate to $\phi(x)$. The parameters $v_{\rm s}$ and $K$ are the spin wave velocity and Luttinger liquid parameter, respectively. Considering the periodicity in $\phi$, the generic perturbation of the form 
\begin{align}
\label{eq:sg}
{\cal H_{\rm {1}}}&= \frac{y_{1}v_{\rm {s}}}{2\pi a^{2}} \int{dx \cos \sqrt{2}\phi}+ \frac{y_{2}v_{\rm {s}}}{2\pi a^{2}} \int{dx \cos 2\sqrt{2}\phi}
\end{align}
is  allowed. Here, $a$ is the short distance cutoff. The higher-order terms $\cos n\sqrt{2}\phi$ ($n \geq 3$) are not considered here, because they are less relevant than ${\cal H_{\rm {1}}}$. Here, $y_1$ forms a spin gap and $y_2$ forms a {\PDN} order. Because $y_1$ is always more relevant than $y_2$,  a  transition from the gapless {\SF} phase to a spin-gap phase should take place before establishing the {\PDN} order, unless $y_1$ identically vanishes. Hence, within this scenario, it is natural to expect a  spin-gap phase between the {\SF} phase and the {\PDN} phase. 

The same scenario is also justified by considering the limiting case of $D_0+\delta D \gg J,\ | D_0-\delta D|$. In this limit,
 the effective Hamiltonian for the spins $\v{S}_{2l}$ is given by (\ref{hameff}) within the second-order perturbation in $J$. 
The antiferromagnetic Ising coupling arises in the third order in $J$. Although other third-order terms also arise, considering that the {\PDN} phase is stabilized on the ordered side of the phase boundary, the spins ${S}^z_{2l}$ and ${S}^z_{2l+2}$ tend to align antiferromagnetically. In addition, the sign of $J_{\rm eff}^{xy}$ can be reversed by the $\pi$-rotation of the spins $\v{S}_{4l}$ around the $z$-axis.  Therefore, the ground state is expected to be similar to that of the $S=2$ XXZ chain with a single-ion anisotropy and a small antiferromagnetic Ising coupling. In this model, assuming continuity to the isotropic case,  a narrow Haldane phase is expected  between the {\SF} phase and the N\'eel phase,\cite{scholl,asha,tonegawa2011} although this has not been confirmed numerically. 
The Haldane phase of the  effective Hamiltonian (\ref{hameff}), which corresponds to the spin-gap phase between the {\SF} and {\PDN} phases in the original  Hamiltonian (\ref{ham0}), is topologically equivalent to the Haldane phase of (\ref{ham0}) 
 near the isotropic point $(D_0,\delta D)=(0,0)$, since both do not have half-integer edge spins under the open boundary condition. From the continuity, this topological equivalence should hold within this phase beyond the limit of applicability of the effective Hamiltonian (\ref{hameff}).  Thus, we may  conclude that the spin-gap phase between the {\SF}  
and {\PDN} phases 
is  topologically equivalent to the Haldane phase near the isotropic point, 
 if the former phase exists.

Motivated by this consideration, we have also estimated the phase boundary, assuming the conventional BKT transition between the {\SF} phase and the spin gap phase using the level spectroscopy with a twisted boundary condition.\cite{nomura-kitazawa}  The finite-size critical point and their extrapolation procedure are also shown in Fig. \ref{xy_neel_bkt_extra} by filled symbols. From the result, we find that the phase boundary estimated in this way almost coincides with that determined by the level crossing of the $\sztot=0$ and $\sztot=1$ excitations. Hence, within the presently available computational resource, we could not reach a definite conclusion regarding the presence of a spin-gap phase between the {\SF} phase and the N\'eel phase, although its presence is naturally expected from 
 the low-energy effective theories.

\section{Summary and Discussion}

The ground-state phases of the  $S=2$ Heisenberg chains with coexisting uniform and alternating single-site anisotropies are investigated by the NED and DMRG methods. The nonmagnetic phase consists of the {\LD}, Haldane, gapless {\SF}, {\PDN}, and conventional N\'eel phases. In addition, partial and quantized ferrimagnetic phases are observed in a wide parameter range.  In contrast to the case of $S=1$, for which the Haldane phase is quite robust,\cite{altd} the Haldane phase in the present model is limited to the close neighborhood of the isotropic point, as in the uniform case.\cite{tonegawa2011} Instead of the Haldane phase, we find a wide gapless {\SF} phase for positive values of $D_0$. 

 The presence of a narrow spin-gap phase between the {\SF} and {\PDN} phases is suggested on the basis of the bosonization argument and mapping onto an effective $S=2$ uniform XXZ chain. We could not, however, confirm the presence of this spin-gap phase within the available numerical data.  As for the  possibility of experimental observation, this intermediate spin-gap phase is so narrow that it would be difficult to detect it, even if a corresponding material is synthesized. Therefore, the  {\SF}-{\PDN} transition would be observed as a direct transition practically. Our argument also applies to 
 the {\SF}-{\PDN} boundary in the spin-alternating chain with competing single-ion anisotropies studied in Ref. \citen{tonemix}, although it would not be numerically detectable. 

As another scenario for the intermediate phase between the conventional gapless {\SF} phase and the {\PDN} phase, a multipolar gapless {\SF} phase with a quasi-long-range order in $(S_i^+)^4$, which is analogous to the XY4 phase in the $S=2$ XXZ chain,\cite{asha,tonegawa2011} may be considered. In this phase, the lowest excitation should have $\sztot=4$. However, we do not find this type of behavior around the {\SF}-{\PDN} phase boundary.

In the partial ferrimagnetic phase, the spontaneous magnetization varies continuously with $D_0$ and $\delta D$, whereas it is locked to a fractional value of the saturated magnetization that satisfies the Oshikawa-Yamanaka-Affleck condition in the quantized ferrimagnetic phase.  The presence of both ferrimagnetic phases in a wide parameter range suggests the realization of magnetized single-chain magnets with a uniform spin magnitude by controlling the environment of each magnetic ion without introducing ferromagnetic interactions.

\acknowledgements

The computation in this work has been performed using the facilities of the Supercomputer Center, Institute for Solid State Physics, University of Tokyo and  Yukawa Institute Computer Facility in Kyoto University. This work is supported by  Grants-in-Aid for Scientific Research (C) (Nos. 25400389 and 21540379) from the Japan Society for the Promotion of Science. The numerical diagonalization program is based on the TITPACK ver.2 coded by H. Nishimori.

\end{document}